# A Modified LSB Technique of Digital Watermarking in Spatial Domain

Nisha Sharma  CITM *           Neelam Malik  CITM*           Kamlesh Sharma  LIMAT

* Manav Rachna International University, Aravali Hills Faridabad

*Abstract* --- **Digital watermarking is a technique of embedding pieces of information into digital data such as text, audio, video, and still images that can be detected or extracted later to show authentication about the data. Watermark is hidden information in the image(s) and is so designed that it does not degrade/distort the quality of the image and still keeps the information. Digital watermarking is basically to protect ownership rights and to control of making illicit copies of digital data. In this paper, we have discussed various watermarking techniques and properties and have proposed a modified LSB technique. We have implemented the proposed technique by following: 2-bits of 8-bit gray image is replaced by luminance part, next 2-bits by red component, next 2-bits by green component and next 2-bits by blue component of 32-bit image using secret key. The advantage is that watermarking capacity has been increased and unaffected by various attacks e.g. zero out LSB bits, cropping etc. Watermark image is imperceptible in resultant image. We have tested this technique on several images and found that it is quite satisfactory. This technique is secured as unauthorized user can not extract the watermarked contents easily from the original image and works well in adverse situations.  We have implemented this technique on platform java 1.5.0.**

*Keywords* **Digital Watermarking, Authentication, Secure, Robust, Copyright Protection**

## I. INTRODUCTION

As multimedia data becomes widespread, such as on the internet, there is a need to address issues related to the security and protection of such data. In recent years security considerations have become very important both in electronic communication and documentation. For protecting multimedia data many cryptography techniques have been developed. But once an authorized user has decoded the message, it can not be restricted for making further illicit copies of such data. Digital Watermarking can be a solution which embeds a secret imperceptible signal into multimedia object so that multimedia object is protected from illegal use. A digital watermark is the distinguishing piece of information that is embedded to data and it is intended to protect multimedia object. It means that it should be very difficult to extract or remove the watermark from the watermarked image. Watermarking of the image could be visible, for example, a background transparent signature, or could be perceptually invisible. A visible watermark acts like a deterrent but may not be acceptable to users in some context. In order to be effective, an invisible watermark should be secure, reliable and resistance to common signal processing operations and intentional attacks.

## II. REQIREMENTS OF IMAGE WATERMARKING

An image watermarking system needs to have at least the following two components:
1. A watermarking embedding system
2. A watermarking extraction (recovery) system.

The watermark embedding system takes input comprising of watermark bit, the image data, and optionally a secret or public key. The output of the watermark embedding system is the watermarked image. The watermark extraction system takes as input an image that possibly contains a watermark and possibly a secret or public key and extracts out the watermark information.

## III. WATERMARKING PROPERTIES

A watermark is designed to permanently reside in the host data. When the ownership of data is in question, the information can be extracted to completely characterize the owner. To achieve maximum protection of intellectual property with watermarked media, several requirements must be satisfied. Of course the properties depend on application such as robust watermarking is required for copyright protection. Among the desirable properties which a watermark should have are:

### A. *Perceptual Transparency*

A digital watermark should not be noticeable to the viewer nor should it degrade the quality of content. The watermark should be imperceptible so as not to affect the viewing experience of the image or the quality of the audio or a video signal.

### B. *Undetectable*

The watermark must be difficult or even impossible to remove by malicious attacker or cracker, at least without obviously degrading the host signal.

### C. *Robustness*

A watermark must be difficult to remove. The attempt to destroy the mark by adding a noise should result in the degradation of the perceptual quality of the host data so as to render it unusable.

### D. *Security*

This is a description of how easy it is to intentionally remove a watermark example by deletion, modification or buying of the watermark in another illicit one.

### E. *Data capacity*

Amount of information that can be stored within the content.

## IV. EXISTING WATERMARKING TECHNIQUES

In general, the embedding techniques can be classified into two categories: spatial domain approach or frequency domain approach.

Spatial domain watermarking technique slightly modifies the pixels of one or two randomly selected subset of an image. The common technique is LSB technique. Its modifications include flipping the lower order bit of each pixel. The LSB technique is in fact simplest technique of watermark insertion. In still images, each pixel of the color image has three components - red, green and blue.

Let us assume we allocate 3 bytes for each pixel. Then, each color has 1 byte, or 8 bits, in which the intensity of that color can be specified on a scale of 0 to 255. For example, the value of a pixel is: $X0 = \{R=254, G=100, B=255\}$. The value of another pixel is: $X1 = \{R=254, G=100, B=254\}$. We have changed all the value of B here. But how much of a difference does it make to the human eye? For the eye, detecting a difference of 1 on a color scale of 256 is almost impossible. Now since each color is stored in a separate byte, the last bit in each byte stores this difference of one. That is, the difference between values 255 and 254, or 127 and 126 is stored in the last bit, called the Least Significant Bit (LSB). Since this difference does not matter much, when we replace the color intensity information in the LSB with watermarking information, the image will still look the same to the naked eye. Thus, for every pixel of 3 bytes (24 bits), we can hide 3 bits of watermarking information, in the LSBs. This uses secret key to choose a random set of bits, and replace them with the watermark. Thus a simple algorithm for this technique would be:

1. Let W be watermarking information

2. for random set of pixel using secret key in the image, Xi

Do Loop: Store the next bit from W in the LSB position of Xi [red] byte, Store the next bit from W in the LSB position of Xi [green] byte, Store the next bit from W in the LSB position of Xi [blue] byte

3. End Loop

To extract watermark information, we would simply need to take all the data in the LSBs of the color bytes and combine them.

Many frequency domains sometimes called the transform domain approaches have been proposed including the discrete cosine transform domain, discrete wavelet transform and discrete Fourier transform. Frequency domain approach has some advantages because most of the signal processing operation can be well characterized in the frequency domain. In frequency domain, values of certain frequencies are altered from their original. The watermark is inserted into coefficients of the transformed image. These frequency alterations are done in the mid frequency components since the low frequency components are very sensitive to distortion and the high frequency components can be removed without significantly affecting the original image quality. The frequency domain watermarking methods are relatively robust to noise, image processing and compression compared with the spatial domain methods. Unfortunately, not too much data can be embedded in frequency domain because the quality of the host image will be distorted significantly.

### V. MODIFIED LSB TECHNIQUE

In this technique, for every pixel of 4 bytes (32 bits), we can hide 8 bits of watermarking information, in the LSBs. In this, 2-bits of watermarking image are replaced with transparency component, 2-bits with red components, 2-bits with green component and 2-bits with blue components. This uses secret key to choose a random set of bits, and replace them with the watermark.

### VI. IMPLEMENTATION

We have implemented proposed modified LSB technique of digital watermarking in java 1.5.0 platform and conducted the experiment of watermarking on various images. The code of modified LSB technique is written in java language. This technique consists of gray level image (watermark image) which is to hide inside the container image i.e. "Sunset's" image. The gray level is 8-bit image and each 2-bits of gray image is embedded into LSB of each 4 components of 32-bits colored image i.e. Sunset image, and the final image is not having any spot able difference. Embedding is done using secret key.

### VII. RESULT AND ANALYSIS

On implementing modified LSB technique of watermarking on original image as shown in fig.(a). The watermarked image as shown in fig.(c) has given the encouraging result.

The visual quality of the image does not change significantly because the watermark bits only change the 2 least significant bits of some pixels with secret key. Hence, the addition of the watermark to an image using this algorithm is quite imperceptible. Using secret key, unauthorized user can not detect the watermark from the image. On the other hand, this algorithm is more robust than LSB technique. Because in LSB technique some attackers can possibly zero out several least significant bits of pixel of the image and hence clear the watermark. This technique has increased the capacity of watermark bits. But in this technique we can detect the presence of watermark after the above attack. But this algorithm will not be robust against JPEG compression because it is performed in the spatial domain and involve least significant bits of the image pixels.

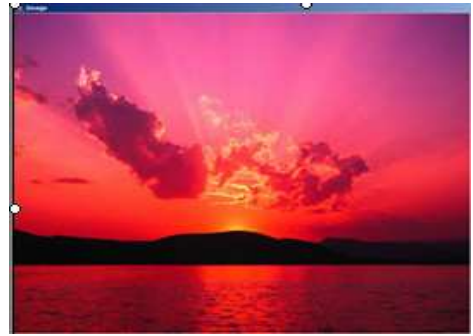

Original Image (a)

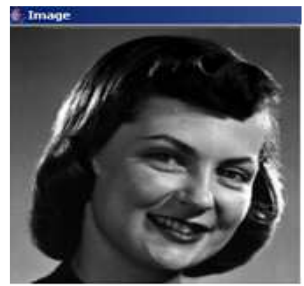

Watermark Image (b)

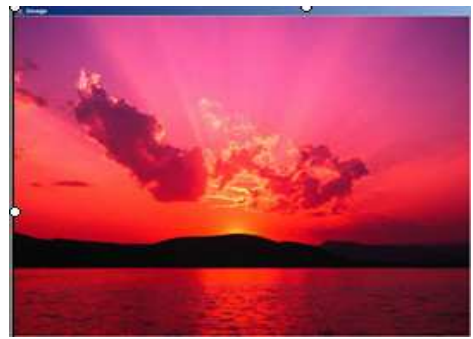

Watermarked Image (c) (Fig. b embedded into Fig. a)

### VIII. CONCLUSION

Thus we conclude that Digital watermarking has importance in securing digital contents from unauthorized user. Modified LSB technique using secret key is more robust than LSB technique. This technique has some advantages over LSB technique. The capacity

Of embedding watermark has been increased and this can detect watermark after some attacks. The result shows that watermarked image seems similar to the original image. It is a secure scheme. Presently we have conducted only spatial domain approach for digital watermarking. In future we will also consider frequency domain approach and we expect that will give more robust and secured watermarked image.

## REFERENCES

[1] Cox, I.J., M.L. Miller, J.M.G. Linnartz and T. Kalker. A Review of Watermarking Principles and practices. pp: 461-482, 1999.

[2] R.B. Wolfgang and E.J. Delp, "A Watermark for Digital Images," *Proc. IEEE Int'l Conf. Image Processing*, vol. 3, pp. 219-222, 1996.

[3] Muhammad Aamir Qureshi and Ran Tao,"A Comprehensive Analysis of Digital Watermarking", I.T. Journal: 471-475, 2006.

[4] Mr. P.D. Khandait, LSB Technique for Secure Data Communication, 2007

[5] Ibrahim Nasir, Ying Weng, Jianmin Jiang," A New Robust Watermarking Scheme for Color Image in Spatial Domain", School of Informatics, University of Bradford, U.K.

[6] M. Kutter and F. A. P. Petitcoals, "A fair benchmark for image watermarking systems", Electronic Imaging '99 Security and Watermarking

of Multimedia Contents , vol. 3657, Sans Jose, CA, USA,25-27 January 1999.

[7] R. Chandramouli and Nasir Memon, "Analysis of LSB Based Image Steganography Techniques", IEEE 2001.

[8] Pori L.Y. and Delina B.," A New Approach in Text Steganography", University of Malaya, Malaysia, 7th Conf. on Applied Computer & Applied Computational Science, China, April 8, 2008.

[9] M. Thaler and H. Winterthur H., "A Java framework for digital image processing "December 22, 2007.

[10] R. K. Sharma and S. Decker: Practical Challenges for Digital Watermarking Application, Digimarc Corporation, USA, 2000.

[11] E. Muharemagic and B.Furht: Multimedia Security: Watermarking Techniques:

Department of Computer Science and Engineering Florida Atlantic University, U.S.A.,E-mail: borko,edin}@cse.fau.edu 1990.